\documentclass[conference]{IEEEtran}
\IEEEoverridecommandlockouts

\usepackage{cite}
\usepackage{amsmath,amssymb,amsfonts}
\usepackage{algorithmic}
\usepackage{graphicx}
\usepackage{textcomp}
\usepackage{xcolor}

\usepackage{float}
\usepackage{color}
\usepackage{amssymb} 
 
\usepackage{verbatim}

\usepackage{caption}
\usepackage{subcaption}
\usepackage{float}
\usepackage{gensymb}

\newtheorem{theorem}{ Theorem}

\newtheorem{definition}{Definition}

\def\BibTeX{{\rm B\kern-.05em{\sc i\kern-.025em b}\kern-.08em
		T\kern-.1667em\lower.7ex\hbox{E}\kern-.125emX}}
\begin{document}
	
	\title{Integrated Sensing and Communication with Nested Array: Beam Pattern  and Performance Analysis\\
	}
	
	\author{Hongqi Min$^{*}$,  Chao Feng$^{*}$, Ruoguang Li$^{\dag}$, and Yong Zeng$^{*\ddag}$ \\
		\IEEEauthorblockA{
			\text{$^{*}$National Mobile Communications Research Laboratory, Southeast University, Nanjing 210096, China} \\
			\text{$^{\dag}$College of Information Science and Engineering, Hohai University, Changzhou 213200, China}\\
			\text{$^{\ddag}$Purple Mountain Laboratories, Nanjing 211111, China}\\
			\text{Email: $\{$minhq, chao$\_$feng, yong$\_$zeng$\}$@seu.edu.cn, ruoguangli@hhu.edu.cn}
			}
	}
	
	\maketitle

	\begin{abstract}
		Towards the upcoming 6G wireless networks, integrated sensing and communication (ISAC) has been identified as one of the typical usage scenarios. To further enhance the performance of ISAC, increasing the number of antennas as well as array aperture is one of the effective approaches.
		However, simply increasing the number of antennas will increase the cost of radio frequency chains and power consumption. 
		To address this issue, in this paper, we consider an uplink ISAC system with nested array deployed at the base station. Nested array is a classic sparse array architecture that is able to enlarge the array aperture without increasing the number of physical antennas. While nested array for wireless sensing has been extensively studied, its potential for ISAC system has not been fully exploited. To fill this gap, in this paper, we provide the beam pattern analysis of nested arrays, and derive the closed-form expressions for the three beam pattern metrics, namely, the main lobe beam width, peak-to-local-minimum ratio, and prominent side lobes height. Extensive simulation results are provided to show that compared with conventional uniform arrays, nested arrays can achieve higher communication performance for densely located users while maintaining its advantage of sensing. 
	\end{abstract}
	
	\section{Introduction}
	\IEEEPARstart 
	{I}{ntegrated} sensing and communication (ISAC) has been identified as one of the main usage scenarios for the sixth generation (6G) wireless networks\cite{9737357,recommendation2023framework}, in which the sensing and communication functionalities will be fully integrated by sharing the hardware platform and wireless resources \cite{9724187}.
	Among the various potential technologies for 6G, extremely large-scale MIMO (XL-MIMO) is a promising candidate that can boost the sensing and communication performance by significantly increasing the number of antennas\cite{10496996,9903389,10545312}.
	Specifically, the augmented number of array antennas and array aperture is able to provide high resolution in the spatial domain, offering a better communication and sensing performance.
	However, for conventional compact MIMO, which sets the inter-antenna spacing to half of the signal wavelength, simply increasing the number of antennas brings higher costs and power consumption \cite{10098681}.
	On the contrary, sparse MIMO, which removes the restriction of half-wavelength antenna spacing,
	is able to enlarge the array aperture without increasing the number of antennas \cite{xinruiLiSparseMIMO}. By deploying the antennas with specific sparse topologies, such as nested arrays, sparse MIMO can achieve enhanced sensing resolution compared to compact MIMO. For the classic nested array, an $N_1$-element compact uniform linear array (ULA) and an $N_2$-element sparse ULA with sparsity $\eta=N_1+1$ are concatenated\cite{10465094}. Besides, sparse array may also provide $\mathcal{O}\left(N^2\right)$ sensing degrees of freedom (DoFs) using only $\mathcal{O}\left(N\right)$ physical antennas when utilizing the second-order statistics of the received signal \cite{5456168}. 
	
	It is worth noting that when the sparse array is implemented in communication systems, the undesired dominating side lobes or grating lobes will be introduced due to the sparse sampling in the spatial domain,
	resulting in severe inter-user interference (IUI)\cite{10465094}.
	Meanwhile, different from sparse array for sensing, it is no longer possible to circumvent these grating lobes by utilizing the second-order statistics of the signal because
	the multiplication of communication symbols will destroy the transmission information.
	However, recent study in \cite{10545312} showed that user grouping and side lobes suppression based on the beamforming optimization can be used to alleviate the impact of grating lobes.
	Furthermore, despite the presence of grating lobes, it is shown in \cite{10465094} that uniform sparse array can obtain better communication performance than the compact counterpart, especially in the hot spot areas when users are densely located. This is because sparse array achieves higher spatial resolution due to narrower main lobe beam, and the probability that users are located at higher-order grating lobes is relatively low, so that the resulting interference can be effectively filtered out.	
	Besides, sparse array is also investigated in the near-field communication \cite{10545312,zhou2024sparse}. 
	Nevertheless, most existing work mainly focuses on sparse array for sensing or communication alone, while its investigation for ISAC is lacking. 
	To be specific, nested arrays have been proven to exhibit satisfactory sensing performance \cite{5456168}, while their potential for wireless communications is largely unexploited. Furthermore, research on their beam pattern synthesis and performance analysis in ISAC systems is completely absent.
	
	Motivated by aforementioned discussions, in this paper, we consider a two-level nested array based ISAC system with $N_1$ elements in the first level and $N_2$ in the second \cite{5456168}.
	Based on this, three critical beam pattern metrics are defined, namely main lobe beam width (BW), peak-to-local-minimum ratio (PLMR), and dominating side lobes height (SLH). 
	Then, closed-form expressions of these metrics are derived for different configuration with parameter $\left(N_1, N_2\right)$.
	It is revealed that the communication gain of nested array is not only dependent on the array architecture $\left(N_1, N_2\right)$, but also on the distribution of users equipments (UEs).
	Last, to compare nested array against the classic compact ULA, simulations based on multi-user uplink ISAC system are conducted, which illustrates that nested array can achieve higher communication data rates for densely located UEs while maintaining its advantage of sensing. On the other hand, nested array can also achieve much better sensing performance in sensing-first architecture while maintaining its advantage of data rates compared to compact ULA. 

	\section{System Model and Signal Processing}
	
	\subsection{System Model}
	\begin{figure}[H]
		\centering
		\includegraphics[width = 7cm]{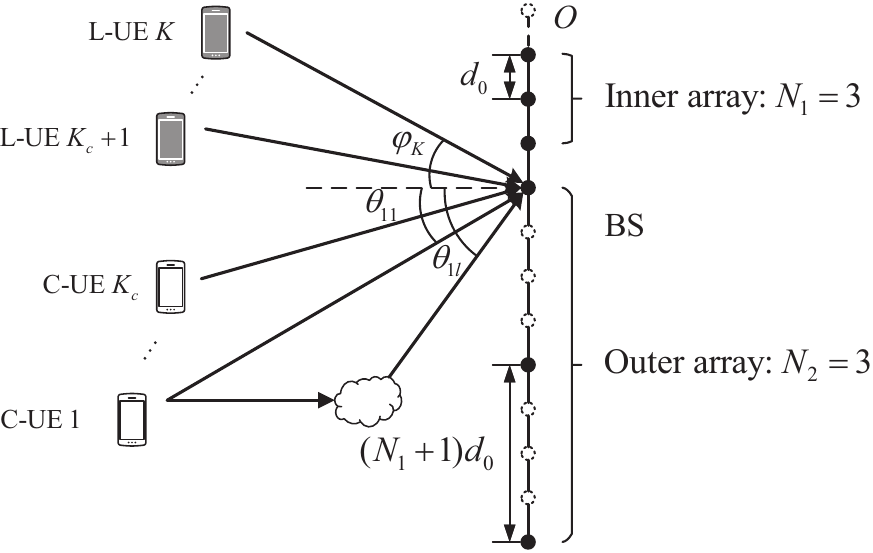}
		\caption{Uplink ISAC system with nested array deployed at the BS.}
		\label{system_1}
	\end{figure}
	As illustrated in Fig. \ref{system_1}, 
	we consider a nested array based uplink ISAC system which consists of an ISAC base station (BS) and $K$ UEs, where the first $K_c$ UEs are communication UE (C-UE) and the rest are localization UE (L-UE). The BS is equipped with a $M$-element sparse array, which needs to perform the uplink communication for C-UEs  and estimate directions of arrival (DoAs) for all L-UEs.
	A nested array on a linear grid is considered at the BS, which consists of an $N_1$-element inner subarray with element spacing $d_0=\lambda/2$ and an $N_2$-element outer subarray with element spacing $\left(N_1+1\right)d_0$, where $M=N_1+N_2$ and $\lambda$ is the wavelength.
	For convenience, a reference point $O$ is set at the origin of the linear grid. Define the distance between the $m$-th antenna and the origin $O$ as $d_m, m=1,2,\cdots,M$.
	Thus, the antenna locations can be expressed as an integer set $\mathcal{D}=\left\{\bar{d}_1, \bar{d}_2, \ldots, \bar{d}_M\right\}$, where $\bar{d}_m=d_m / d_0$ \cite{5456168}.
	\begin{definition}
		\label{fenchao}
		A nested array is defined in terms of the integer parameter pair $\left(N_1, N_2\right)$, which is the union of a compact ULA and a sparse ULA, i.e., $\mathcal{D}=\mathcal{D}_{in}\cup \mathcal{D}_{ou}$\cite{10368470}:
		\begin{equation}
			\resizebox{0.9\hsize}{!}{$
				\begin{aligned}
				\text{compact ULA } \mathcal{D}_{in}:& \left\{1,2, \ldots, N_1\right\},\\
				\text{Sparse ULA } \mathcal{D}_{ou}:&\left\{N_1+1, 2 (N_1+1), \ldots, N_2 (N_1+1)\right\}.
			\end{aligned}$}
		\end{equation}
	\end{definition}
	Note that the nested array can degenerate to compact ULA when $(N_1, N_2)=(0,M) \text{ or } (M-1,1)\text{ or }(M,0)$.
	
	The channel vector between C-UE $k$ and the BS can be expressed as 
	\begin{equation}
		\label{minhongguang}
		\mathbf{h}_k=\sum\nolimits_{i=1}^{L_k} \beta_{k i} \mathbf{a}\left(\theta_{k i}\right), k=1,2,\cdots, K_c,
	\end{equation}
	where $L_k$ is the number of multi-paths of C-UE $k$ and $\beta_{ki}$ denotes the complex-valued channel gain of the $i$-th path. Besides, $\mathbf{a}\left(\theta_{k i}\right)=\left[1,e^{j\pi\left(\bar{d}_2-1\right)\sin\theta_{ki}},...,e^{j\pi\left(\bar{d}_M-1\right)\sin\theta_{ki}}\right]^{T}$ is the array steering vector of C-UE $k$ with $\theta_{ki}$ representing its angle-of-arrival (AoA) of the $i$-th path. 
	Furthermore, assume that the non-line-of-sight (NLoS) links between L-UEs and BS are negligible, so that
	the channel between the $q$th L-UE and BS can be given by $\mathbf{h}_q=\beta_q\mathbf{a}\left(\varphi_{q}\right), q=K_c+1,\cdots, K$, with $\beta_q$ and $\varphi_{q}$ representing the channel gain and DoA of L-UE $q$, respectively. 	
	
	Thus, the received signal at the BS can be expressed as
	\begin{equation}
		\label{zhangguangjun}
		\mathbf{y}=\sum\nolimits_{i=1}^{K} \mathbf{h}_i \sqrt{P_i} x_i +\mathbf{n},
	\end{equation}
	where $x_i$ and $P_i$ denote the independent information-bearing symbol and transmit power of UE $i$, respectively;
	$\mathbf{n} \sim \mathcal{C} \mathcal{N}\left(0, \sigma^2 \mathbf{I}_M\right)$ is the additive white Gaussian noise (AWGN). 
	
	\subsection{Communication Signal Processing}
	To detect the communication signal for C-UE $k$, a linear receive beamforming vector $\mathbf{v}_k \in \mathbb{C}^{M \times 1}$ with $\left\|\mathbf{v}_k\right\|=1$ is used. Thus, the resulting signal is
	\begin{equation}
		\resizebox{0.9\hsize}{!}{$\begin{aligned}
			y_k=\mathbf{v}_k^H \mathbf{y}=\mathbf{v}_k^H \mathbf{h}_k \sqrt{P_k} x_k+\mathbf{v}_k^H \sum_{i=1, i \neq k}^{K} \mathbf{h}_i \sqrt{P_i} x_i
			+\mathbf{v}_k^H \mathbf{n}.\\
		\end{aligned}$}
	\end{equation}	
	
	The signal-to-interference-plus-noise ratio (SINR) of C-UE $k$ is given by
	\begin{equation}
		\begin{aligned}
			\gamma_k=\frac{P_k\left|\mathbf{v}_k^H \mathbf{h}_k\right|^2}{\sum_{i=1, i \neq k}^K P_i\left|\mathbf{v}_k^H \mathbf{h}_i\right|^2
			+\sigma^2}.
		\end{aligned}
	\end{equation}
	Therefore, the achievable rate of C-UE $k$ can be expressed as 
	\begin{equation}
		\label{wangsihan}
		R_k=\log _2(1+\gamma_k) .
	\end{equation}

	For the special LoS-dominating case, we have $L_k=1$. In this case, the channel vector can reduce to
		\begin{equation}
			\resizebox{0.9\hsize}{!}{$
				\mathbf{h}_k=\beta_{k} \mathbf{a}\left(\theta_{k}\right)
			=\beta_k\left[1, e^{j \pi \left(\bar{d}_2-1\right) \sin \theta_k}, \ldots, e^{j \pi \left(\bar{d}_M-1\right) \sin \theta_k}\right]^T,$}
		\end{equation}
	where $\beta_k$ and $\theta_k $ are the channel gain and AoA of C-UE $k$, respectively.
	By using the maximal-ratio combining (MRC) beamforming vector $\mathbf{v}_k=\frac{\mathbf{h}_k}{\|\mathbf{h}_k\|}$, the corresponding SINR can be simplified as
	\begin{equation}\label{bf_methods}
		\resizebox{0.9\hsize}{!}{$\begin{aligned}
			\gamma_{k} 
		 = \frac{P_k\left|\beta_k\right|^2M}
			{
				\sum\limits_{i=1, i \neq k}^{K} P_i M \left|\beta_i\right|^2 \left|\frac{\mathbf{h}_k^H \mathbf{h}_i}{\left\|\mathbf{h}_i\right\|\left\|\mathbf{h}_k\right\|}\right|^2 
				+\sigma^2
			}
			= \frac{\bar{P}_k M}
			{M \sum\limits_{i=1, i \neq k}^{K} \bar{P}_i \rho_{k i}
				+1}
		\end{aligned}$},
	\end{equation}
	where $\bar{P}_k \triangleq \frac{\left|\beta_k\right|^2 P_k}{\sigma^2}$ is the receive SNR
	and $\rho_{ki}=\frac{\left|\mathbf{h}_k^H \mathbf{h}_i\right|^2}{\left\|\mathbf{h}_k\right\|^2\left\|\mathbf{h}_i\right\|^2}$ denote the channel's squared-correlation coefficient
	between UE $k$ and $i$.

	\subsection{Localization Signal Processing}
	For localization, (\ref{zhangguangjun}) can be rewritten as
	\begin{equation}
		\mathbf{y}=\mathbf{A}\mathbf{s}+\mathbf{n}
		=\begin{bmatrix}\mathbf{A}_c &  \mathbf{A}_l\end{bmatrix}
		\begin{bmatrix}
			\mathbf{s}_c\\
			\mathbf{s}_l
		\end{bmatrix}
		+\mathbf{n},
		\label{fushen}
	\end{equation}
	where the steering matrix $\mathbf{A}=\begin{bmatrix}\mathbf{A}_c &  \mathbf{A}_l\end{bmatrix} $, in which $\mathbf{A}_l=$ $[\mathbf{a}(\varphi_{K_c+1}) ,\cdots,\mathbf{a}(\varphi_{K})]\in\mathbb{C}^{M\times \left(K-K_c\right)}$ and
	$\mathbf{A}_c=\left[ \mathbf{A}_{1}\boldsymbol{\beta}_{1},\cdots,\mathbf{A}_{K_c}\boldsymbol{\beta}_{K_c}\right]\in\mathbb{C}^{M\times K_c} $ with $\mathbf{A}_k=\left[ \mathbf{a}(\theta_{k1}),\cdots,\mathbf{a}(\theta_{kL_k})\right] $ as well as $\boldsymbol{\beta}_{k}=\left[\beta_{k1},\cdots,\beta_{kL_k}\right]^T$.
	Besides, $\mathbf{s}_c=\left[\sqrt{P_1} x_1, \cdots, \sqrt{ P_{K_c}}x_{K_c}\right]^T\in\mathbb{C}^{K_c\times 1}$ 
	and $\mathbf{s}_l=\left[\beta_{K_c+1}\sqrt{P_{K_c+1}} x_{K_c+1}, \cdots,\right.$ $\left.\beta_{K}\sqrt{P_{K}} x_{K}\right]^T\in \mathbb{C}^{\left(K-K_c\right)\times1}$ are signals from C-UEs and L-UEs, respectively. 
	Besides, it is assumed that the communication symbols $\mathbf{s}_c$ have been detected based on communication signal processing, and we substract the communication signals in ($\ref{fushen}$), leaving only the localization signals as
	\begin{equation} \label{youxiaohu}
			\mathbf{y}_l=\mathbf{A}_l\mathbf{s}_l+\mathbf{n}.
	\end{equation}
		
	Note that classic DoA estimation cannot be directly performed on (\ref{youxiaohu}) due to the non-uniform spatial sampling by the nested array. Specifically, the beamspace based algorithms will suffer from angle ambiguity. Besides, the super resolution subspace based algorithms, such as ESPRIT\cite{32276}, require the vandermond property of $\mathbf{A}$. 
	To this end, a virtual receive signal model can be obtained to address this issue and exploit the potential of nested array, which is based on the second-order statistics of the received signal\cite{5456168}. 
	First, the covariance matrix of the received signal is given by
	\begin{equation}
		\begin{aligned}
				\mathbf{R}  &=\mathbb{E}\left[\mathbf{y}_l\mathbf{y}_l^H\right] =\mathbf{A}_{l} \mathbf{R}_{l} \mathbf{A}_{l}^H
				+\sigma_n^2 \mathbf{I}_M, \\
			\end{aligned}
		\end{equation}
	where 
	$\mathbf{R}_{l}=\text{diag}\{ |\beta_{K_c+1}|^2P_{K_c+1},\cdots,|\beta_{K}|^2P_{K}\} $ is the covariance matrix of $\mathbf{s}_l$.
	Then, we vectorize $\mathbf{R}$ to obtain
	\begin{equation}
		\label{zRxx}
		\begin{aligned}
			\mathbf{z} & =\operatorname{vec}\left(\mathbf{R}\right)
			=\left(\mathbf{A}_l^* \odot \mathbf{A}_l\right) \mathbf{r}+\sigma_n^2 \overrightarrow{\mathbf{1}},
		\end{aligned}
	\end{equation}
	where  
	$\mathbf{r}=\left[|\beta_{K_c+1}|^2P_{K_c+1},\cdots,|\beta_{K}|^2P_{K}\right]^T$  
	can be regarded as the equivalent receive signal, and $\sigma_n^2\overrightarrow{\mathbf{1}}=\sigma_n^2
	\begin{bmatrix}
		\mathbf{e}_1^T, \mathbf{e}_2^T,\cdots ,\mathbf{e}_M^T
	\end{bmatrix}^T$ is the equivalent receive noise where $\mathbf{e}_i$ is a zeros vector except 1 at the $i$th position. 
	Meanwhile, the conjugate and KR product $\odot$ operations behave like the difference of the positions of the physical antennas, which forms a vitual array whose antennas locate at $\mathcal{D}_z=\{d_i-d_j,i,j =1,2,\cdots,M\}$.
	After performing vitual array antennas selection and sorting, an equivalent steering matrix whose receive antennas locate at the distinct values of $\mathcal{D}_z$ can be obtained.
	It has been proved that $\mathcal{D}_z$ can form a compact ULA with at most $\frac{M^{2}+2M-2}{2}$ antennas by using only $M$ physical antennas when $M$ is even and $N_1=N_2=\frac{M}{2}$ \cite{5456168}. Therefore, the DoFs and array aperture are significantly augmented, rendering it possible to improve the sensing performance.
	Note that (\ref{zRxx}) becomes a single snapshot signal, thereby spatial smoothing needs to be adopted to recover the rank of the covariance matrix of $\mathbf{z}$. Based on this, classic DoA estimation algorithms can be performed to obtain the DoAs of the L-UEs.

	\section{Beam Pattern of Nested Array}	\label{sec3}
	From (\ref{bf_methods}), it can be observed that the communication data rate of C-UE $k$ is mainly influenced by $\rho_{k i}$, which depends on the beam pattern of nested array. Beam pattern reflects the array gain for a designed beam with respect to the spatial angle and it also significantly affects the sensing accuracy\cite{7386582}. 
	Therefore, in order to figure out whether nested arrays can outperform the standard compact arrays for ISAC systems, the beam pattern is analyzed firstly.	
	Consider the nested array $\left(N_1, N_2\right)$ defined in \emph{Definition \ref{fenchao}},
	the beam pattern can be expressed as
	\begin{equation}
		\label{qianhua}
		\resizebox{0.89\hsize}{!}{$\begin{aligned}
				\rho_{k i}&
				= \left|\frac{1}{M} \mathbf{a}^H\left(\theta_k\right)\mathbf{a}\left(\theta_i\right)\right|^2
				= \left|\frac{1}{M} \sum_{m=0}^{M-1} a_m^*\left(\theta_i\right) a_m\left(\theta_k\right)\right|^2\\
				&=\frac{1}{M^2}\left|
				\frac{\sin (\frac{\pi}{2} N_1\bar{d}_{in}\Delta_{ki})}{\sin (\frac{\pi}{2}\bar{d}_{in}\Delta_{ki})}
				+e^{j\frac{\pi}{2}N_2\bar{d}_{ou}\Delta_{ki}}
				\frac{\sin(\frac{\pi}{2} N_2\bar{d}_{ou}\Delta_{ki})}{\sin(\frac{\pi}{2} \bar{d}_{ou}\Delta_{ki})}\right|^2\\
				&\triangleq G(\Delta_{ki}),\\
			\end{aligned}$}
	\end{equation}
	where $\bar{d}_{in}=1$ and $\bar{d}_{ou}=N_1+1$ are the element spacing of inner and outer subarrays on the uniform grid, respectively. Besides, $\Delta_{ki}\triangleq\sin \theta_k-\sin \theta_i \in[-2,2]$ is the spatial angle difference and $a_m\left(\theta_i\right)$ is the $m$th value of $\mathbf{a}\left(\theta_i\right)$. 
	Owning to the non-uniform architecture of physical array, the corresponding beam pattern of nested array $G(\Delta)$ has grating lobes and dominating side lobes.
	For ease of reading, the subscript of $\Delta_{ki}$ is omitted in the following.
	
	To gain some insights, the beam pattern in (\ref{qianhua}) can be simplified as
	\begin{equation}\label{yanhongping}
		\resizebox{0.89\hsize}{!}{$\begin{aligned}
			G(\Delta)&=\frac{1}{M^2}\left|\lambda(\Delta)\right|^2
			=\frac{1}{M^2}\left|f(\Delta)+e^{j\Phi(\Delta)}g(\Delta)\right|^2\\
			&=\frac{1}{M^2}\left[f^2(\Delta)+g^2(\Delta)+2f(\Delta)g(\Delta)\cos(\Phi(\Delta))\right],\\
		\end{aligned}$}
	\end{equation}
	where $\lambda(\Delta)=f(\Delta)+e^{j\Phi(\Delta)}g(\Delta)$, $f(\Delta)=\frac{\sin (\frac{\pi}{2} N_1\bar{d}_{in}\Delta)}{\sin (\frac{\pi}{2}\bar{d}_{in}\Delta)}$, $g(\Delta)=\frac{\sin(\frac{\pi}{2} N_2\bar{d}_{ou}\Delta)}{\sin(\frac{\pi}{2} \bar{d}_{ou}\Delta)}$ and $\Phi(\Delta)=\frac{\pi}{2}N_2\bar{d}_{ou}\Delta$. It is found that the main lobe of $G$ can be approximately regarded as the combination of the main lobes of $f$ and $g$ except a complex coefficient $e^{j\Phi}$. 
	In order to obtain more main lobe characteristics, 
	the first local minimum point (FLMP) $\Delta_{min}$ is defined.
	\begin{definition}
		\label{zhouzhiwen} The FLMP of $G(\Delta)$ is defined as the smallest local minimum point on the positive semi-axis.
		\begin{equation}
			\Delta_{min}= \min\{\Delta \mid \Delta\in\{\Delta_{i}^{loc}\}_{i=1}^{n}\},
		\end{equation}
		 where $\{\Delta_{i}^{loc}\}_{i=1}^{n}$ are the $n$ local minimum points of $G(\Delta)$.
	\end{definition}

	\subsection{Main lobe Beam Width (BW)}
	The main lobe width of nested array is defined as $\text{BW}=2\Delta_{min}$.
	Although the FLMP of $G(\Delta)$ can not be solved in closed-form, the first null points of $f$, $g$ and $\Phi$ can be easily obtained.
	For $f(\Delta)$, when $\frac{\pi}{2} N_1\bar{d}_{in}\Delta=\pi$, yielding
	$\Delta_1=\frac{2}{N_1\bar{d}_{in}}=\frac{2}{N_1}$;
	for $g(\Delta)$, when $\frac{\pi}{2} N_2 \bar{d}_{ou}\Delta=\pi$, yielding
	$\Delta_2=\frac{2}{N_2\bar{d}_{ou}}=\frac{2}{(N_1+1)N_2}$;
	for $\cos(\Phi(\Delta))$, when $\frac{\pi}{2} N_2 \bar{d}_{ou}\Delta=\frac{\pi}{2}$, yielding
	$\Delta_3=\frac{1}{N_2\bar{d}_{ou}}=\frac{1}{(N_1+1)N_2}$.
	Note that we have $\Delta_1=\frac{\left(N_1+1\right)N_2}{N_1}\Delta_2$, $\Delta_2 =2\Delta_3 $ and $\Delta_3  < \Delta_2  < \Delta_1$.
	Then, it can be proved that $\Delta_{min}$ has lower and upper bounds. 

	\begin{theorem}
		\label{wuyulin}
		The FLMP $\Delta_{min}$ is bounded by
		\begin{enumerate}
			\item When $ 2\leq N_2\leq N_{th}$,
			\begin{equation}
				\frac{2(N_2-1)}{(N_1+1)N_2} \leq\Delta_{min} \leq \frac{2}{N_1+1},
			\end{equation}
			\item When $ N_{th}< N_2\leq N_{ap}$,
			\begin{equation}
				\frac{1}{(N_1+1)N_2} \leq\Delta_{min} \leq \frac{2}{(N_1+1)N_2} , 
			\end{equation}
			\item When $ N_2> N_{ap}$,
			\begin{equation}
				\Delta_{int}\leq\Delta_{min} \leq\frac{2}{(N_1+1)N_2} ,
			\end{equation}
		\end{enumerate}
		where $\Delta_{int}$ is a unique solution to the equation $\cos(\Phi)=\frac{g(\Phi)}{2f(0)}$ for $\Phi\in[\frac{\pi}{2},\pi]$. Besides, $N_{th}$ and $N_{ap}$ are two threshold values of $N_2$ for a given $N_1$, where
		\begin{equation*}
			\resizebox{0.85\hsize}{!}{$
				N_{th}= \left\{
				\begin{aligned}
					&1,& N_1<7 \\
					&\max\{N_2 \mid G^{\prime}\left(\Phi\right) < 0,  0 \leq \Phi \leq \pi  \}, &N_1\geq 7\\
				\end{aligned}
				\right.$}
		\end{equation*}
		and $N_{ap} \approx \lfloor \sqrt{\frac{10N_1^2}{N_1+1}} \rfloor$.
	\end{theorem}
	
	\begin{IEEEproof}
		See Appendix  \ref{liruoguang}.
	\end{IEEEproof}
	
	Note that when the outer subarray is much smaller than the inner subarray, i.e., $N_2\leq N_{th}\ll N_1$, the BW of nested array mainly depends on the inner subarray in the asymptotic case, i.e., $\text{BW}\rightarrow  \text{BW}_{in}=2\Delta_1$ for $N_1\rightarrow+\infty$.
	By contrast, when the outer subarray is large enough, i.e., $ N_2\gg N_{ap}$, the BW is determined by the outer subarray $\text{BW}\rightarrow  \text{BW}_{ou}=2\Delta_2$ for $N_2\rightarrow+\infty$. 
	
	\subsection{Peak-to-Local-Minimum Ratio (PLMR)}
	PLMR often characterizes the SNR that UE can achieve in the main lobe, which can be defined as follows. 
	\begin{definition}
		\label{yuzhi} PLMR is the ratio of $G(0)$ to the beam gain of the FLMP, i.e., 
		$
			\text{PLMR }= \frac{G\left(0\right)}{G\left(\Delta_{min}\right)},
		$
		where $G\left(0\right)=(N_1+N_2)^2/M^2=1$.
	\end{definition}
	
	It is worth noting that the FLMP $\Delta_{min}$ does not have a closed-form expression. Thus, lower bound of PLMR is used to approximate the actual value.
	\begin{theorem}
		\label{fengyingqi}
		PLMR of a nested array $\left(N_1, N_2\right)$ is  
		\begin{enumerate}
			\item When $ 2\leq N_2\leq N_{th}$, 
			$
				\text{PLMR }\geq \max\{\frac{1}{P_4}, \frac{1}{P_5}\},
			$
			\item When $ N_{th}\leq N_2\leq N_{ap}$, 
			$
				\text{PLMR }\geq \max\{ \frac{1}{P_{3}}, \frac{1}{P_{2}}\},
			$
			\item When $ N_2\geq N_{ap}$,
			$
				\text{PLMR }\geq \max\{ \frac{1}{P_{int}}, \frac{1}{P_{2}}\},
			$
		\end{enumerate}
		where $P_{1}= G(\Delta_1 )=\frac{1}{M^2}g^2(\Delta_1 )$, 
		$P_{2} = G(\Delta_2 )=\frac{1}{M^2}f^2(\Delta_2 )$, 
		$P_{3} = G(\Delta_3 )=\frac{1}{M^2}\left(f^2(\Delta_3 )+g^2(\Delta_3 )\right)$, 
		$P_{4}=G\left(\left(N_2-1\right)\Delta_2\right)=\frac{1}{M^2}f^2\left(\left(N_2-1\right)\Delta_2\right)$, 
		$P_{5} =G\left(N_2\Delta_2 \right)=\frac{1}{M^2}\left(f^2\left(N_2\Delta_2 \right)+N_2^2-2N_2f\left(N_2\Delta_{2}\right)\right)$,
		$P_{int}= G(\Delta_{int})$.
	\end{theorem}
	
	\subsection{Prominent Side Lobe Height (SLH)}
	Considering the more general case that all the UEs are randomly distributed in the direction of $\left[-\pi/2, \pi/2\right]$, the interfering users may appear not only in the main lobe and the FLMP directions, but also in side lobe directions. Therefore, the direction and height of the dominating side lobes are also an important metric reflecting the IUI of communication. 
	
	\begin{theorem}
		\label{qianxiufeng}
		Given $N_2> N_{ap}$, the dominating side lobes appear at 
		$
			\Delta_{\mathrm{s, n}} \approx \frac{2n}{N_1+1}, n = \pm1, \pm2,\cdots, \pm  N_1, 
		$
		and these lobes have the similar height, which is
		$
			\text{SLH } \approx \frac{\left(N_2-1\right)^2}{M^2}.
		$
	\end{theorem}
	\begin{IEEEproof}
		The position of the outer subarray's grating lobes can be obtained by letting $\frac{\pi}{2} \bar{d}_{ou}\Delta=n\pi, n = \pm1, \pm2,\cdots , \pm  N_1$, and thus the $n$-th grating lobe will appear at
		$
		\Delta_{g,n}=\frac{2n}{N_1+1}, n = \pm1, \pm2,\cdots, \pm  N_1.
		$		
		The height of these grating lobes are  $\frac{\left(N_2-1\right)^2}{M^2}$. It can be observed that the $N_2>N_{ap}$ is able to make the outer subarray's grating lobes prominent in the beam pattern of the nested array, and these dominating side lobes in $G$ approximately appear at $\Delta_{g,n}, n = \pm1, \pm2,\cdots, \pm  N_1$ with height $\frac{\left(N_2-1\right)^2}{M^2}$.
	\end{IEEEproof}
	
	Otherwise, when $N_2 \leq N_{ap}$, the beam pattern of nested array depends mainly on the inner subarray because the amplitude of the outer subarray is too small compared with the inner one, submerging the grating lobes of the outer subarray in the beam pattern of nested array. Therefore, the SLH of this case is similar to the inner subarray and the details are omitted here. 
	
\begin{figure}[H]
		\centering
		\includegraphics[width = 7cm]{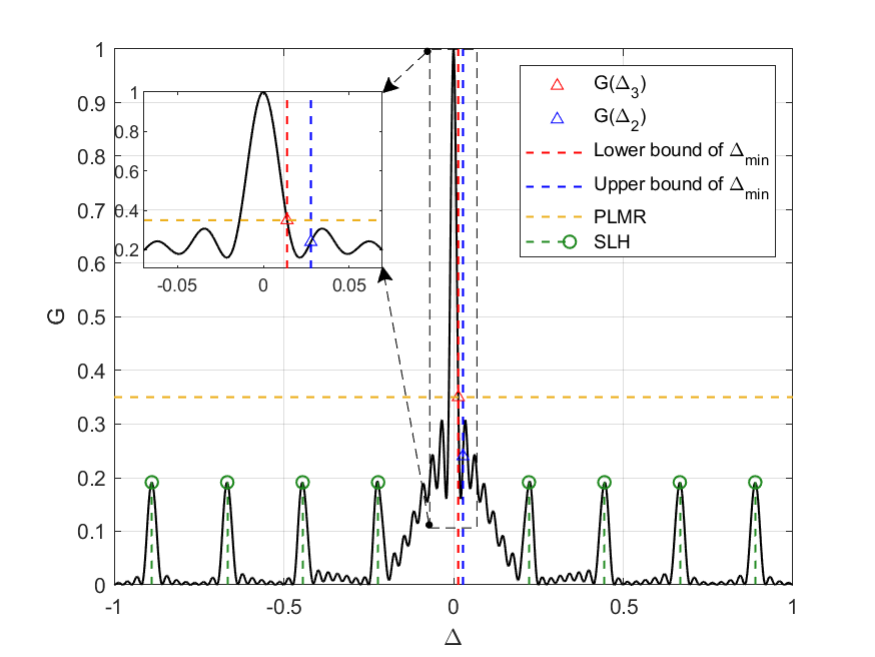}
		\vspace*{-0.3cm}
		\caption{Beam pattern of nested array with $\left(N_1,N_2\right)=\left(8,8\right)$.}
		\label{beampattern_na}
	\end{figure}

	As shown in Fig. \ref{beampattern_na}, the main metrics of nested array, i.e., FLMP, PLMR and SLH, can be efficiently obtained, where $\left(N_1,N_2\right)=\left(8,8\right)$. 	
	After that, flexible design of nested arrays, side lobes suppression based on the beamforming optimization and even user grouping can be realized to mitigate IUI introduced by these side lobes \cite{xinruiLiSparseMIMO}.  	

	\section{Numerical Results}
	Unless otherwise stated, we set $\bar{P}_k=20 \mathrm{~dB}$. For the multi-path channel in (\ref{minhongguang}), we use the ``one-ring" model, with $L_k=10$ multi-paths, $R=5 \mathrm{~m}$ denoting the radius of each ring, and $r=40 \mathrm{~m}$ denoting the range of the center of the ring \cite{995514}. The Rician factor is set to be $20 \mathrm{~dB}$. The UEs are uniformly distributed in direction $\left[-\theta_{max},\theta_{max} \right]$.
	
\begin{figure*}[htbp]
		\centering
		\begin{minipage}{0.325\linewidth}
			\centering
			\includegraphics[width=1\linewidth]{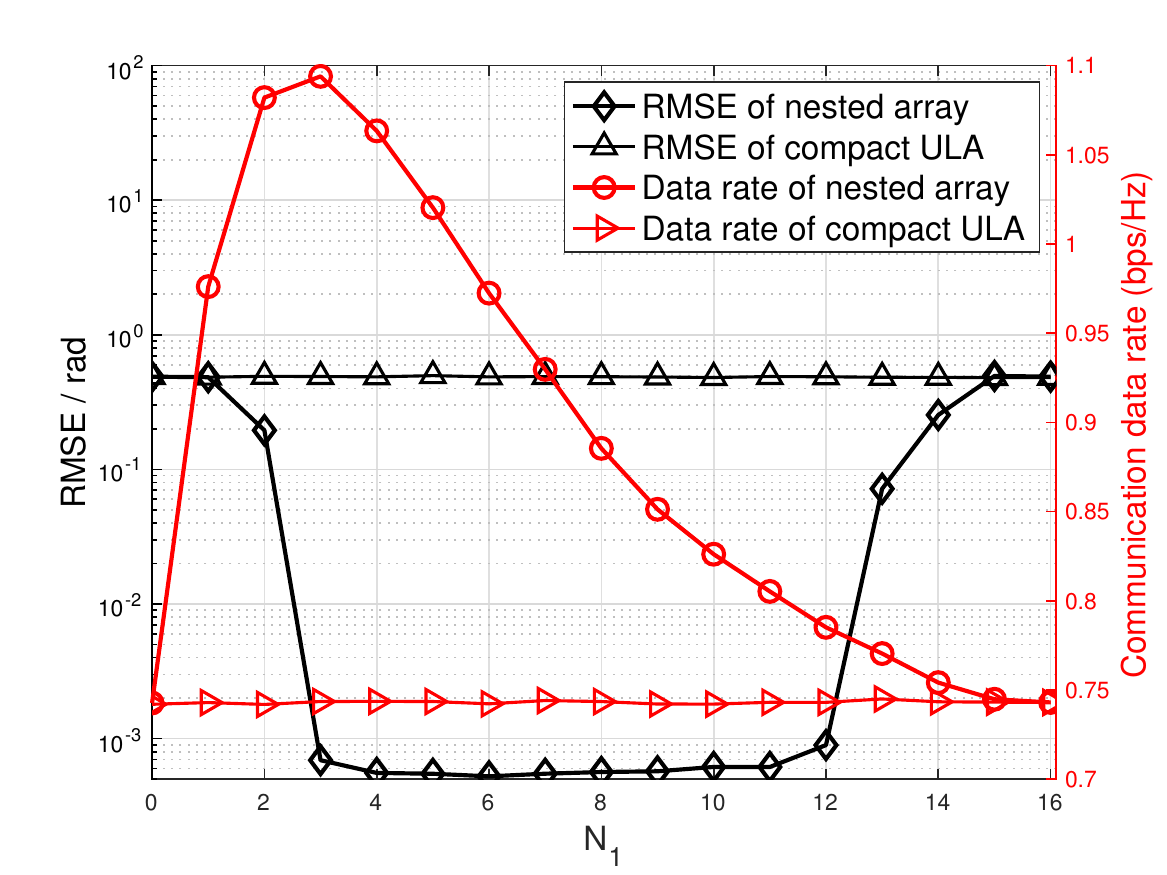}
			\caption{RMSE and data rate v.s. $N_1$.}
			\label{fig3}
		\end{minipage}
		\begin{minipage}{0.325\linewidth}
			\centering
			\includegraphics[width=1\linewidth]{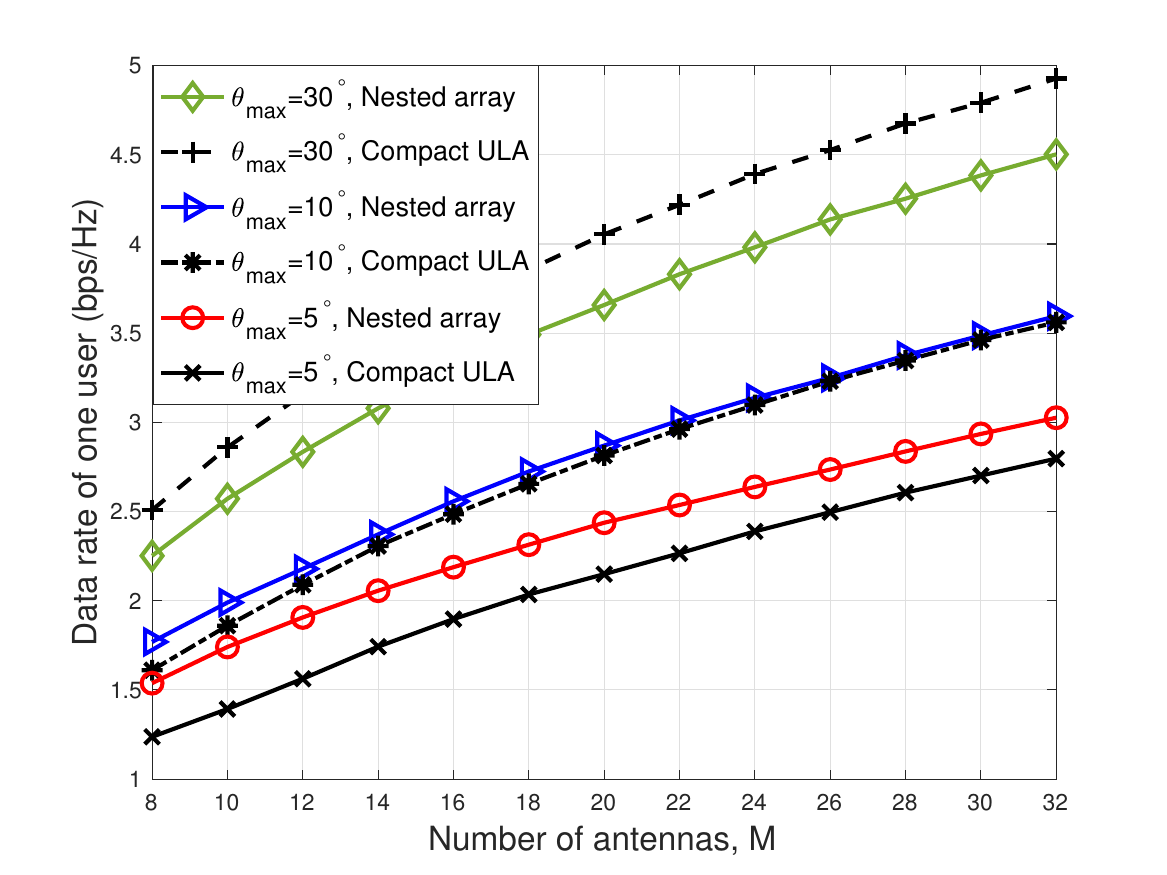}
			\caption{Data rate v.s. $M$.}
			\label{fig4}
		\end{minipage}
		\begin{minipage}{0.325\linewidth}
			\centering
			\includegraphics[width=1\linewidth]{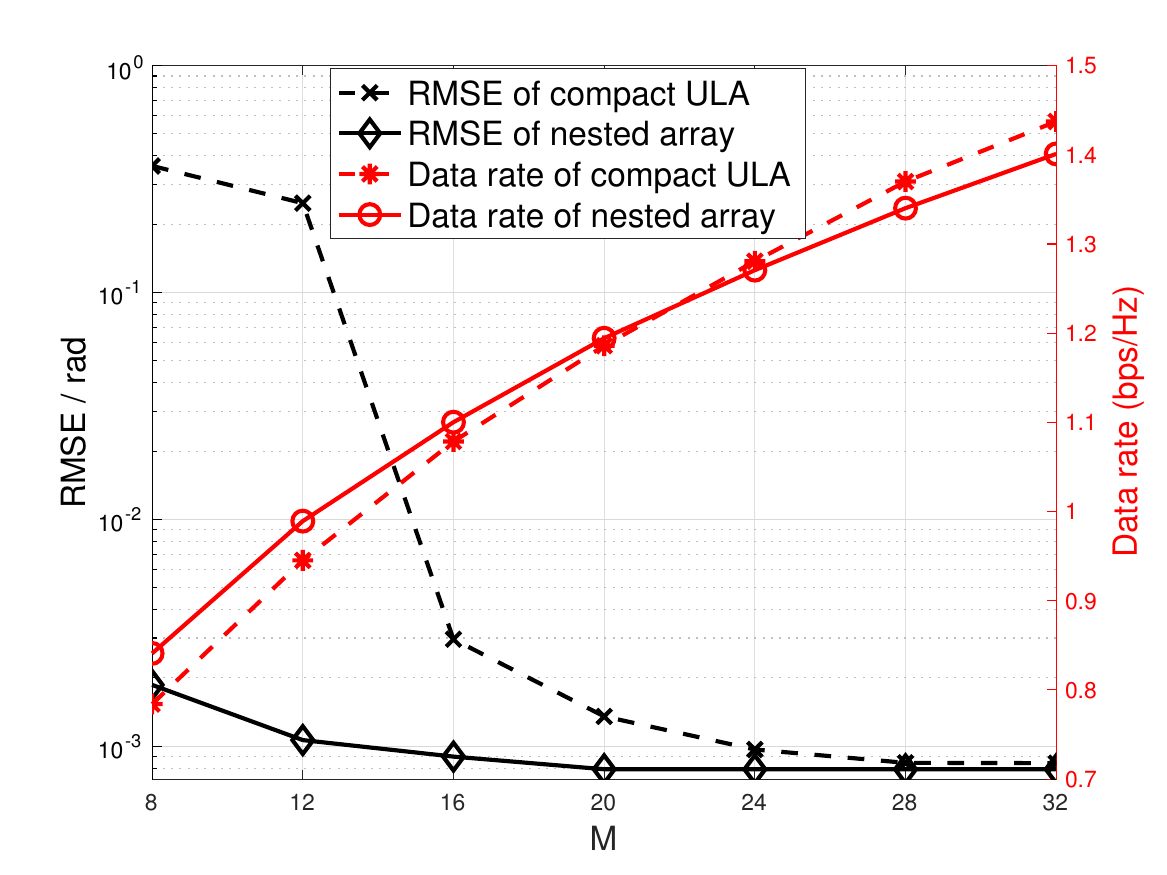}
			\caption{RMSE and data rate v.s. $M$.}
			\label{fig5}
		\end{minipage}
		\vspace*{-0.3cm}
	\end{figure*}

	Fig. \ref{fig3} shows that, for UEs uniformly distributed in the main lobe area of compact ULA, i.e., $\theta_{max}=3.58^{\circ}$, the nested arrays always have better communications data rates while maintaining its advantage of sensing than compact ULA with the same number of antennas $M=16$. $N_1$ is set to increase from 0 to $M$, $N_2=M-N_1$, $K=7$ and $L=1$. On the one hand, it is easy to see that the data rates of nested array and compact ULA are identical at $N_1=0$, $N_1=M-1$ and $N_1=M$ because the nested arrays degenerate to compact ULA at these cases. 
	On the other hand, the optimal data rate occurs at $N_1=3$ because it suppress the IUI efficiently 
	with balanced BW, PLMR and SLH configurations for densely distributed UEs. 

	Fig. \ref{fig4} shows that the communications data rate gain of nested arrays is not only influenced by the array architecture but also by the distribution of UEs.
	Firstly, it is obvious that the data rates of both nested array and compact ULA increase as $M$ increasing.
	Secondly, when UEs are densely distributed in hot spot areas when $\theta_{max}=5^{\circ}$, nested arrays obtain obvious better data rates because the small main lobe BW providing high spatial resolution to mitigate IUI effectively. Last but not least, this benefits may decrease when $\theta_{max}$ increases to $10^{\circ}$ and even deteriorate when $\theta_{max}$ further increases to $30^{\circ}$. This is because more dominating side lobes of nested array will contribute to the IUI as $\theta_{max}$ increase. 
	
	Fig. \ref{fig5} indicates that the nested arrays based on sensing-first architecture, that is $N_1=N_2=\frac{M}{2}$ with the largest virtual array for sensing, can achieve much better sensing performance compared to compact ULA with the same $M$ while maintaining its advantage of data rates. Simulation parameters are $\theta_{max}=18^\circ$ and $M$ increases from 8 to 32. 
	It is easy to see that data rates and RMSEs of both nested arrays and compact ULAs are getting better as $M$ increases. 
	Meanwhile, there exists a intersection point for the two data rate lines, this is because one new side lobe occur in the UEs area $\theta\in[-18^\circ, 18^\circ]$ as $M$ increases, which is derived as $\Delta_{g,n}=\frac{2n}{N_1+1}$ in Section \ref{sec3}. 
 	
	\section{Conclusions}
 	In this paper, beam pattern analysis of nested arrays for ISAC system is provided and the proposed three beam pattern metrics are studied. Thus the main characteristics of nested arrays for ISAC can be obtained effectively. Furthermore, we apply the nested arrays to an integrated sensing and communication system.
 	Simulation shows that nested arrays can offer better communication data rates when UEs are densely distributed while maintaining its advantage of sensing. On the other hand, nested array can also obtain much better sensing performance in sensing-first architecture while maintaining its advantage of data rates compared to compact ULA. 
	
	{\appendices
		\section{Proof of Theorem \ref{wuyulin}} \label{liruoguang}
		
		\subsection{$N_2>N_{ap}$}
		For (\ref{yanhongping}), the FLMP of $G$ is equivalent to the minimum length of the vector $\vec{\lambda} =\vec{f} +e^{j\Phi}\vec{g}$ in the complex plane as shown in Fig. \ref{xujingran}(a).
		$\Phi$ is the angle between the vector $\vec{g}e^{j\Phi}$ and the horizontal axis. As $\Delta$ increases, $\Phi$ also increases, and the vector $\vec{g}e^{j\Phi}$ rotates counterclockwise. Then, the position pointed by $\vec{\lambda}$ will form a trajectory, which is the solid blue line.
		
		Firstly, based on the facts that the first null points of $f$, $g$ and $cos(\Phi(\Delta))$ follow $\Delta_3  < \Delta_2  < \Delta_1$, which indicates $f$, $g$ and $cos(\Phi(\Delta))$ decrease monotonically for $\Delta\in [0,\Delta_3]$, so $G$ also decreases monotonically in this interval and $\Delta_{min}\geq \Delta_3$. 
		For simplicity, we assume that $|\vec{f}|\approx f(0)\triangleq f_c, \text{ for } \Delta \in [0,\Delta_2]$ since $f(\Delta)$ decreases slowly compared with $g(\Delta)$. Therefore, the trajectory can be approximated by the red dashed line. Note that the red trajectory can perfectly estimate the blue one, demonstrating the rationality of this approximation.		
		
		Since $\Delta_3$ corresponds to $\Phi=\frac{\pi}{2}$ and we have $\Delta_2\Delta_3\perp O\Delta_2$, a contour line with radius $f_c$, which is centered at the origin $O$ will be tangent to $\Delta_2\Delta_3$ at $\Delta_2$.
		Meanwhile, the red trajectory and the horizontal axis are also tangent at $\Delta_2$. Consequently, there must be an intersection between the trajectory and the contour line for $\Delta\in\left(\Delta_3,\Delta_2\right)$, which is denoted by $\Delta_{int}$.
		Due to $|O\Delta_{int}|=|O\Delta_{2}|$, we have $G(\Delta_{int})=G(\Delta_{2})\approx \frac{1}{M}f_c^2$ and a local minimum point will exist in the interval $[\Delta_{int},\Delta_{2}]$.
		Therefore, we can obtain
		$
			\Delta_{3}\leq \Delta_{int}\leq \Delta_{min}\leq \Delta_2.
		$
		
		In addition, $\Delta$ can be specifically expressed as $G(\Phi)=\frac{1}{M}f_c^2$, which is
		$
			f_c^2+g^2(\Phi)+2f_cg(\Phi)cos(\Phi)=f_c^2,
		$
		where $g(\Phi)=\frac{\sin(\Phi)}{\sin(\Phi/N_2)}$. Solving it yields
		$
			g(\Phi)=0 \text{ or }\cos(\Phi)=-\frac{g(\Phi)}{2f_c}
		$.		
		The former corresponds to $\Phi=\pi$, i,e, $\Delta=\Delta_2$. For the latter equation, because $cos (\Phi) $ decreases monotonically for $\Phi$ from $0$ to $-1$, and $-\frac{g(\Phi)}{2f_c}$ increases monotonically from $-\frac{g(\Phi)|_{\Phi=\frac{\pi}{2}}}{2f_c}<0$ to $0$. Therefore, there must be a unique solution $\Phi_{int}\in[\frac{\pi}{2},\pi]$, and then we can obtain $\Delta_{int}=\frac{2\Phi_{int}}{\pi N_2\bar{d}_{ou}}$. 
		
		\subsection{$N_{th}< N_2\leq N_{ap}$}
		With $N_2$ decreasing, the variation of $f$ in the interval $\left[0,\Delta_2\right]$ is more considerable than that of $g$. As a result, the assumption that $f$ can be regarded as a constant in the case $N_2>N_{ap}$ will become no longer reasonable. Specifically, $f$ decreases from $N_1$ to 0 in the interval $[0,\Delta_1]$, while $g$ decreases from $N_2$ to 0 in the interval $[0,\Delta_2]$, where $\Delta_1=\frac{\left(N_1+1\right)N_2}{N_1}\Delta_2$. Thus, the average decreasing velocity of $f$ and $g$ can be obtained as $\frac{N_1^2}{N_2(N_1+1)\Delta_2}$ and $\frac{N_2}{\Delta_2}$, respectively. Furthermore, when $\frac{10N_1^2}{N_2(N_1+1)\Delta_2} < \frac{N_2}{\Delta_2}$, the variation of $f$ can be neglected,
		thus yielding $N_{ap}=\lfloor\sqrt{\frac{10N_1^2}{N_1+1}}\rfloor$.
		
		As shown in Fig. \ref{xujingran}(b), the red trajectory cannot match well with the blue trajectory due to $N_2\leq N_{ap}$. 
		However, $\Delta_{min}\geq \Delta_3$ is still valid, even though it is not a tight lower bound. 
		Besides, $|\vec{\lambda}|$ may own an a increasing interval in $\Delta\in\left(\Delta_3,\Delta_2\right)$ since $N_2 > N_{th}$. 
		Thus, we can have $\Delta_3 \leq\Delta_{min} \leq \Delta_2$.
\begin{figure}[H]	
	\centering
	\begin{subfigure}{0.9\linewidth}
		\centering
		\includegraphics[width=0.7\linewidth]{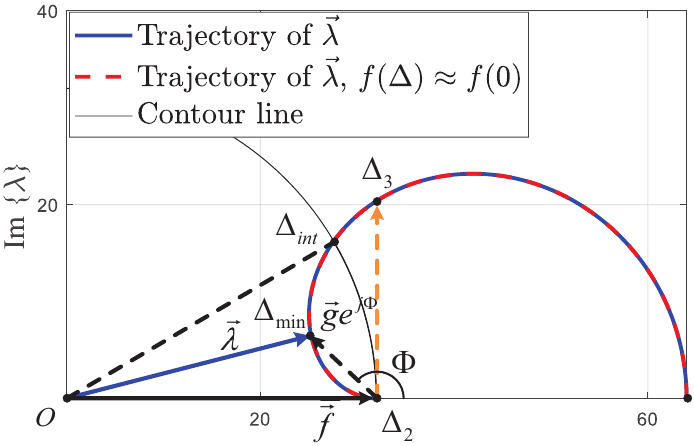}
		\caption{Nested array with $\left(N_1,N_2\right)=\left(32,32\right)$.}				
	\end{subfigure}
	\centering
	\begin{subfigure}{0.9\linewidth}
		\centering
		\includegraphics[width=0.7\linewidth]{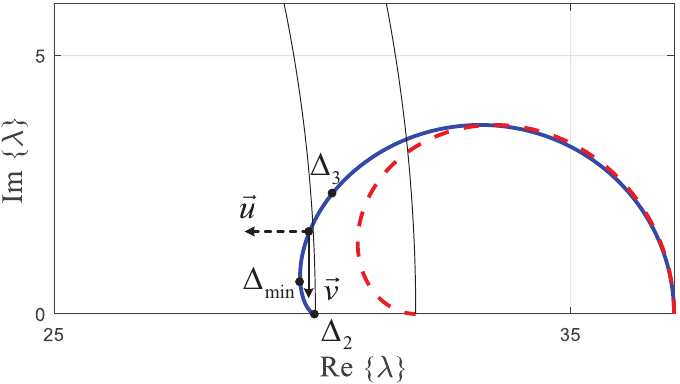}
		\caption{Nested array with $\left(N_1,N_2\right)=\left(32,5\right)$.}
	\end{subfigure}	
	\caption{Trajectory of $\vec{\lambda}$ for nested array with  $\left(N_1,N_2\right)=\left(32,32\right)$ and $\left(32,5\right)$.}	
	\label{xujingran}
\end{figure}
		
		\subsection{$2\leq N_2\leq N_{th}$}
		As $N_2$ further reduces, the local minimum point will no longer appear in the interval $\left[0,\Delta_2\right]$. Specifically, when the direction of the $\vec{\lambda}$ trajectory is decomposed to two orthogonal directions, one is $\vec{u}$ which is paralleling to ${O\Delta_2}$, and the other is $\vec{v}$ perpendicular to $\vec{u}$ as shown in Fig. \ref{xujingran}(b). It is easy to know that the FLMP will exist in the interval $\left[\Delta_3, \Delta_2\right]$ when an inversion of $\vec{u}$ happens with $\Phi$ increasing from 0 to $\pi$. 
		In fact, $|\vec{u}|$ depends on the decreasing velocity of $|\vec{f}|$ as well as $|\vec{g}e^{j\Phi}|$. Specifically, $|\vec{u}|$ will point to the left when $\frac{d|\vec{f}|}{d\Phi}\gg\frac{d|\vec{g}|}{d\Phi}$, yielding decreasing of $|\vec{\lambda}|$, otherwise increasing. Consequently, 
		it can be predicted that when $N_2$ is less than a threshold $N_{th}$, $|\vec{u}|$ will always point to the left on the interval $\Delta\in\left(\Delta_3,\Delta_{2}\right)$, so that there is no local minimum point.
		
		However, $N_{th}$ cannot be solved in closed-form due to the non-linearity of $G$. A useful searching method can be adopted to get $N_{th}$. Firstly, 
		let $N_2$ take values from 1 to $N_1$ and then numerically calculate the derivative of G for each $N_2$.
		Once the numerical derivative array occurs positive value, $N_{th}=N_2-1$. It is noted that $N_{th}=1$ when $N_1<7$. In addition, $N_{th}$ is very small compared to $N_1$, such as $N_{th}=4$ for $N_1=32$ and $N_{th}=11$ for $N_1=512$. Thus, $N_{th}$ can be expressed as
		$N_{th}=\max\{N_2 \mid G^{\prime}\left(\Phi\right) < 0,  0 \leq \Phi \leq \pi  \}$ for $N_1\geq 7$.
		
		Next, it is can be understand that $G$ decreases monotonically for $\Delta\in\left(\Delta_2,\left(N_2-1\right)\Delta_2\right)$, where $\left(N_2-1\right)\Delta_2<\Delta_1$. 
		The reason is that $f$ and $g$ are both similar to the $sinc$ function, which has a high main lobe and significantly lower side lobes, and this interval is corresponding to the main lobe of $f$ while to the side lobes of $g$.
		So the increase-decrease characteristics of $|\vec{\lambda}|$ depends mainly on $f^2$ since $N_1\gg N_{th}\geq N_2$. Therefore, $\Delta_{min}\geq \left(N_2-1\right)\Delta_2$.	
		
		Last, it can be proved that $\frac{dG}{d\Delta}\big|_{\Delta=N_2\Delta_2}
		=\frac{2df}{d\Delta}\big|_{\Delta=N_2\Delta_2}
		\cdot \left(f\left(N_2\Delta_2\right)+g\left(N_2\Delta_2\right)\cos\left(N_2\pi\right)\right)>0$ since $\frac{2dg}{d\Delta}\big|_{\Delta=N_2\Delta_2}=\frac{2d\cos\left({\Phi}\right)}{d\Delta}\big|_{\Delta=N_2\Delta_2}=0$. Meanwhile, $G$ is a continuously differentiable function, so $G$ increases in the neighborhood of $\Delta=N_2\Delta_{2}$. Therefore, $\Delta_{min}\leq N_2\Delta_2$.	The proof is as follows: firstly, it is apparent that $\frac{df}{d\Delta}\big|_{\Delta=N_2\Delta_2}<0$, $f\left(N_2\Delta_2\right)>0$ and $g\left(N_2\Delta_2\right)\cos\left(N_2\pi\right)<0$. Secondly, $ \lim\limits_{N_1\rightarrow +\infty}N_2\Delta_2=\Delta_1$, yielding $\lim\limits_{N_1\rightarrow +\infty}f\left(N_2\Delta_2\right)=0$. Besides, $|g\left(N_2\Delta_2\right)\cos\left(N_2\pi\right)|=N_2$, yielding $f\left(N_2\Delta\right)+g\left(N_2\Delta\right)\cos\left(N_2\pi\right)<0$.
		 Finally, the proof of $\left(N_2-1\right)\Delta_2\leq\Delta_{min}\leq N_2\Delta_2$ is finished. 
	}

	\bibliographystyle{IEEEtran}
	\bibliography{reference}

\begin{thebibliography}{10}
\providecommand{\url}[1]{#1}
\csname url@samestyle\endcsname
\providecommand{\newblock}{\relax}
\providecommand{\bibinfo}[2]{#2}
\providecommand{\BIBentrySTDinterwordspacing}{\spaceskip=0pt\relax}
\providecommand{\BIBentryALTinterwordstretchfactor}{4}
\providecommand{\BIBentryALTinterwordspacing}{\spaceskip=\fontdimen2\font plus
\BIBentryALTinterwordstretchfactor\fontdimen3\font minus
  \fontdimen4\font\relax}
\providecommand{\BIBforeignlanguage}[2]{{%
\expandafter\ifx\csname l@#1\endcsname\relax
\typeout{** WARNING: IEEEtran.bst: No hyphenation pattern has been}%
\typeout{** loaded for the language `#1'. Using the pattern for}%
\typeout{** the default language instead.}%
\else
\language=\csname l@#1\endcsname
\fi
#2}}
\providecommand{\BIBdecl}{\relax}
\BIBdecl

\bibitem{9737357}
F.~Liu, Y.~Cui, C.~Masouros, J.~Xu, T.~X. Han, Y.~C. Eldar, and S.~Buzzi,
  ``Integrated sensing and communications: Toward dual-functional wireless
  networks for {6G} and beyond,'' \emph{{IEEE} J. Select. Areas Commun.},
  vol.~40, no.~6, pp. 1728--1767, Jun. 2022.

\bibitem{recommendation2023framework}
ITU-R, ``Framework and overall objectives of the future development of {IMT}
  for 2030 and beyond,'' Draft New Recommendation, Jun. 2023.

\bibitem{9724187}
Z.~Xiao and Y.~Zeng, ``Waveform design and performance analysis for full-duplex
  integrated sensing and communication,'' \emph{{IEEE} J. Select. Areas
  Commun.}, vol.~40, no.~6, pp. 1823--1837, Jun. 2022.

\bibitem{10496996}
H.~Lu, Y.~Zeng, C.~You, Y.~Han, J.~Zhang, Z.~Wang, Z.~Dong, S.~Jin, C.-X. Wang,
  T.~Jiang, X.~You, and R.~Zhang, ``A tutorial on near-field {XL-MIMO}
  communications towards {6G},'' \emph{IEEE Commun. Surv. Tuts.}, pp. 1--1,
  2024.

\bibitem{9903389}
M.~Cui, Z.~Wu, Y.~Lu, X.~Wei, and L.~Dai, ``Near-field {MIMO} communications
  for {6G}: Fundamentals, challenges, potentials, and future directions,''
  \emph{IEEE Commun. Mag.}, vol.~61, no.~1, pp. 40--46, Jun. 2023.

\bibitem{10545312}
X.~Li, Z.~Dong, Y.~Zeng, S.~Jin, and R.~Zhang, ``Multi-user modular {XL-MIMO}
  communications: Near-field beam focusing pattern and user grouping,''
  \emph{IEEE Trans. Wireless Commun.}, pp. 1--1, Aug. 2024.

\bibitem{10098681}
Z.~Wang, J.~Zhang, H.~Du, W.~E.~I. Sha, B.~Ai, D.~Niyato, and M.~Debbah,
  ``Extremely large-scale {MIMO}: Fundamentals, challenges, solutions, and
  future directions,'' \emph{IEEE Wireless Commun.}, pp. 1--9, Jun. 2023.

\bibitem{xinruiLiSparseMIMO}
X.~Li, H.~Min, Y.~Zeng, S.~Jin, L.~Dai, Y.~Yuan, and R.~Zhang, ``Sparse {MIMO}
  for {ISAC}: New opportunities and challenges,'' \emph{arXiv preprint
  arXiv:2406.12270}, 2024.

\bibitem{10465094}
H.~Wang and Y.~Zeng, ``Can sparse arrays outperform collocated arrays for
  future wireless communications?'' in \emph{2023 IEEE Globecom Workshops (GC
  Wkshps)}, 2023, pp. 667--672.

\bibitem{5456168}
P.~Pal and P.~P. Vaidyanathan, ``Nested arrays: A novel approach to array
  processing with enhanced degrees of freedom,'' \emph{IEEE Trans. Signal
  Processing}, vol.~58, no.~8, pp. 4167--4181, Aug. 2010.

\bibitem{zhou2024sparse}
C.~Zhou, C.~You, H.~Zhang, L.~Chen, and S.~Shi, ``Sparse array enabled
  near-field communications: Beam pattern analysis and hybrid beamforming
  design,'' \emph{arXiv preprint arXiv:2401.05690}, 2024.

\bibitem{10368470}
C.~Zhou, Y.~Gu, Y.~D. Zhang, and Z.~Shi, ``Sparse array interpolation for
  direction‐of‐arrival estimation,'' in \emph{Sparse Arrays for Radar,
  Sonar, and Communications}, Hoboken, NJ, USA: Wiley, 2024, pp. 41--74.

\bibitem{32276}
R.~Roy and T.~Kailath, ``{ESPRIT}-estimation of signal parameters via
  rotational invariance techniques,'' \emph{IEEE Trans. Acoust., Speech, Signal
  Processing}, vol.~37, no.~7, pp. 984--995, Jul. 1989.

\bibitem{7386582}
S.~Kwak, J.~Chun, D.~Park, Y.~K. Ko, and B.~L. Cho, ``Asymmetric sum and
  difference beam pattern synthesis with a common weight vector,'' \emph{IEEE
  Antennas Wireless Propagat. Lett.}, vol.~15, pp. 1622--1625, Jun. 2016.

\bibitem{995514}
A.~Abdi and M.~Kaveh, ``A space-time correlation model for multielement antenna
  systems in mobile fading channels,'' \emph{IEEE J. Select. Areas Commun.},
  vol.~20, no.~3, pp. 550--560, Apr. 2002.

\end{thebibliography}
	
\end{document}